\documentclass{article}

\usepackage{times}
\usepackage{graphicx} 
\usepackage{subfigure}

\usepackage{natbib}

\usepackage{algorithm}
\usepackage{algorithmic}

\usepackage{hyperref}

\newcommand{\remove}[1]{}

\usepackage[accepted]{icml2017}

\usepackage{graphicx}
\usepackage{wrapfig}

\icmltitlerunning{Sharkzor}

\begin{document}

\twocolumn[
\icmltitle{Sharkzor: Interactive Deep Learning for Image Triage, Sort and Summary}



\icmlsetsymbol{equal}{*}

\begin{icmlauthorlist}
\icmlauthor{Meg Pirrung}{pnnl-r}
\icmlauthor{Nathan Hilliard}{pnnl-s}
\icmlauthor{Art{\"e}m Yankov}{pnnl-s}
\icmlauthor{Nancy O'Brien}{pnnl-r}
\icmlauthor{Paul Weidert}{pnnl-r}
\icmlauthor{Courtney D Corley}{pnnl-r}
\icmlauthor{Nathan O Hodas}{pnnl-r}
\end{icmlauthorlist}

\icmlaffiliation{pnnl-r}{Pacific Northwest National Laboratory, Richland, WA, USA}
\icmlaffiliation{pnnl-s}{Pacific Northwest National Laboratory, Seattle, WA, USA}

\icmlcorrespondingauthor{Nathan O Hodas}{nathan.hodas@pnnl.gov}
\icmlcorrespondingauthor{Courtney D Corley}{court@pnnl.gov}

\icmlkeywords{deep learning, machine learning, user centered design, HCI}

\vskip 0.3in
]



\printAffiliationsAndNotice{}  

\begin{abstract}
Sharkzor is a web application for machine-learning assisted image sort and summary. Deep learning algorithms are leveraged to infer, augment, and automate the user's mental model. Initially, images uploaded by the user are spread out on a canvas. The user then interacts with the images to impute their mental model into the application’s algorithmic underpinnings. Methods of interaction within Sharkzor's user interface and user experience support three primary user tasks; \textbf{triage}, \textbf{organize} and \textbf{automate}. The user triages the large pile of overlapping images by moving images of interest into proximity. The user then organizes said images into meaningful groups. After interacting with the images and groups, deep learning helps to automate the user's interactions.  The loop of interaction, automation, and response by the user allows the system to quickly make sense of large amounts of data.
\end{abstract}

\section{Introduction}
On their own, state-of-the-art deep learning systems are typically capable of achieving accuracies between 80\% and 99\% for tasks such as image classification, given sufficient training data. While satisfactory for research purposes, these accuracies may not suffice for using deep learning in everyday tools. For example, 99\% accuracy in a self-driving car would lead to unacceptable deaths. In the case of self-driving cars, whenever the deep learning system has low confidence in its next course of action, operation of the vehicle would be turned over to a human. Facebook's facial recognition provides another example of how a human in the loop supplements a deep learning system. While Facebook's deep learning-based facial recognition system can achieve some 97\% accuracy~\cite{Taigman}, Facebook will still call upon its users to help verify a given face whenever the deep learning classifier has low confidence in its prediction. Thus, human in the loop centric design enables deep learning's ubiquity in every-day applications.

The image organization task requires human in the loop design -- only the human analyzing the photos knows the optimal organization of images for their particular task. Machines may be good at grouping images in ways that make sense to machines, but not to humans, unless explicitly primed with some notion of what the algorithm should group by (eg. color or shape)~\cite{hodas2016}. Sharkzor supports human-machine interaction by using physical proximity on the canvas and grouping as information for the system. The essence of Sharkzor is the power to embed an image’s high dimensionality into a two-dimensional space. In the Sharkzor system we aim to keep the human at the center of the task, rather than as a simple and tedious algorithm trainer.

Users move images around on a  2D web-based canvas, arranging things freeform or into groups.  Upon request, the Sharkzor system  repositions images or regroups them to reflect its assessment of the user's mental model.  The user may then refine the system's suggestions.  In this way, the user may retrain the system, and the user may understand Sharkzor's  deep learning models by assessing the system's organization of images in comparison to the user's mental model. Thus, Sharkzor enables dynamic human in the loop machine learning  to enable image sort and summary.

\subsection{Related Work}
Some of the earliest and contemporarily relevant research dealing with  human-automation interaction can be traced back to \cite{Parasuraman} and \cite{Cummings}. In~\cite{Cummings}, research concentrated on Navy weapons operators who had to synthesize instant messaging data from multiple sources to make supervisory decisions on how to control Tomahawk missiles. \remove{More recently, researchers at Facebook have used a human-in the loop framework to train bots to communicate with humans \cite{Li}. In this framework, reinforcement learning is used to improve the bot's question-answering ability from feedback a teacher gives following the bot's generated responses.} In \cite{Yu}, an iterative loop consisting of humans and deep learning is used to generate a dataset containing some one million labeled images of ten scene categories and twenty object categories. Finally, in \cite{Yi} the researchers were able to utilize a human-in the loop system to train a convolutional neural network to segment foreground moving objects in surveillance videos. The performance of the classifier rivaled that of humans while reducing the manual labor involved in ground truthing the videos by up to forty times. Recently, \cite{hodas2016} presented a system for using the two-dimensional arrangements of images to capture mental model of the user and position images accordingly. However,this system required the user to touch every image.\footnote{See \cite{hodas2016} for a review of other systems that automate 2D arrangements of images.}   Sharkzor leverages deep learning to learn the user's mental model and apply it to images it has not seen before.

\section{Human-In the Loop Design}
%
%
Though many applications of machine learning to human in the loop tasks utilize the human as a fall-back when the algorithm is unsure, we instead focus on using machine learning algorithms to augment the user's own organizational methodology. Our user-centered approach, which focuses primarily on optimizing the user and their task, also support users with different and unexpected workflows.

The Sharkzor system works more similarly to a recommender system like Netflix or Amazon. Sharkzor requires few training examples (as few as 2, initially), and instead integrates the training process seamlessly into the user's normal workflow. Sharkzor also provides feedback and insight into the machine learning process by way of confidence visualization and image heatmap overlay.

\subsection{User-Centered Design}
The User-Centered Design Cycle \cite{Norman:1986:UCS:576915} describes a method for designing and developing systems that focus, as the name suggests, on the user. We utilized this method for researching, designing, developing  the Sharkzor software.

During the research phase of the cycle, we inventoried features \remove{and supported tasks} of other machine learning image organization tools \cite{smpe} and general image organization tools \cite{revisit, pivot}. We then refined and elaborated on the supported tasks to support the intended Sharkzor user. We created a task taxonomy, grouping the lower level tasks and relevant interactions into three primary system tasks: \textbf{triage}, \textbf{organize} and \textbf{automate}.

The Sharkzor interface is designed to be a flexible system that allows users to complete their tasks in non-predetermined ways. We provide a system where how and when to utilize the machine learning augmentation is up to the user throughout their workflow.

\subsubsection{User Tasks}
We identified user tasks and interactions throughout our research phase, then prioritized and organized the interactions using our task taxonomy based around the main triage, organize and automate task families.

\begin{figure}[t]
  \centering
  \includegraphics[width=0.47\textwidth]{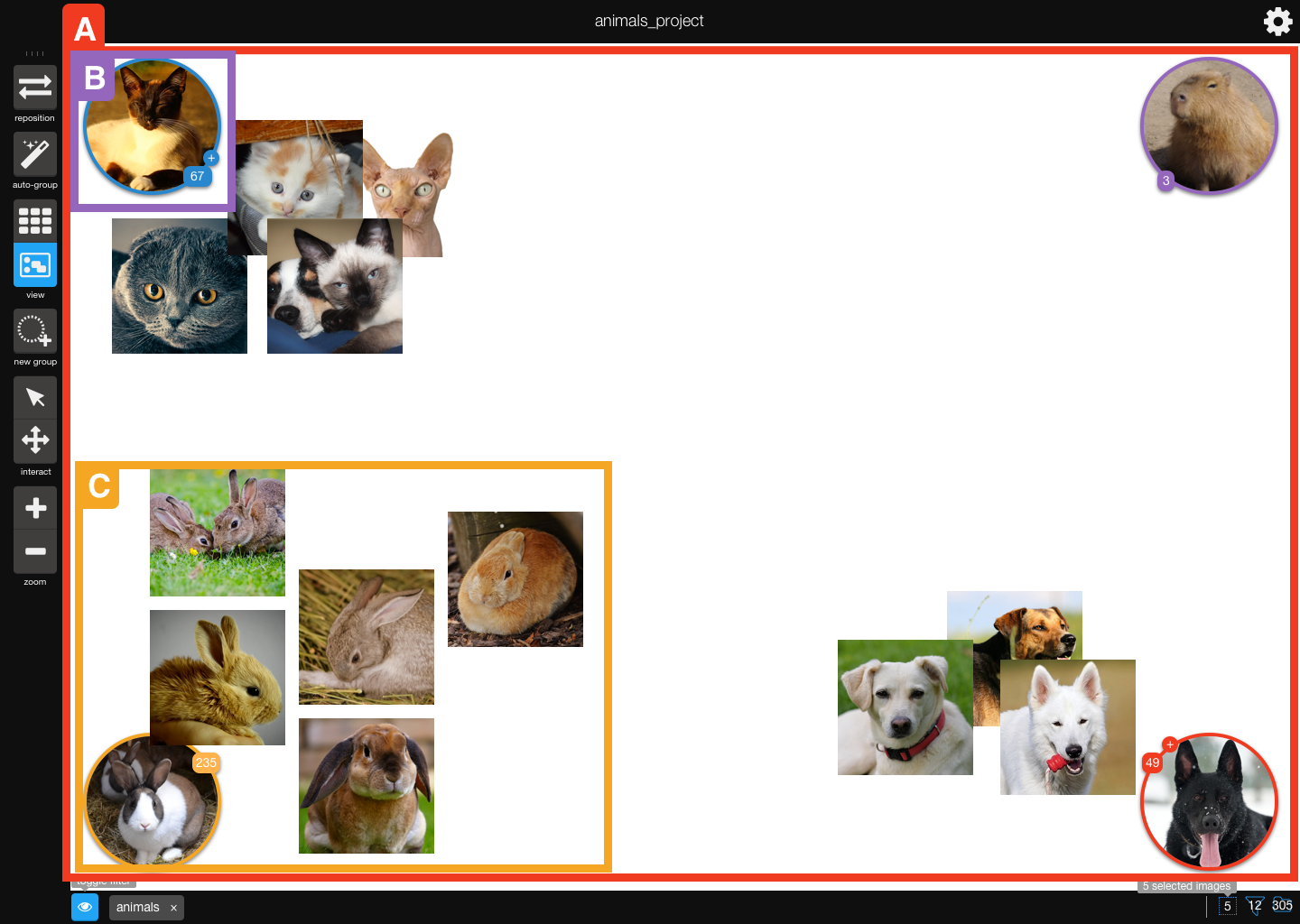}
   \caption{Sharkzor interface canvas view. (A) The Sharkzor canvas on which the user can organize images and groups. (B) A user created group with images that have been auto-grouped via machine learning functionality, indicated by the plus badge. (C) Images that have been repositioned near visually similar images after the user has created a group and moved images into its proximity.}
   \label{fig:1}
\end{figure}

\begin{figure}[h]
  \centering
  \includegraphics[width=0.47\textwidth]{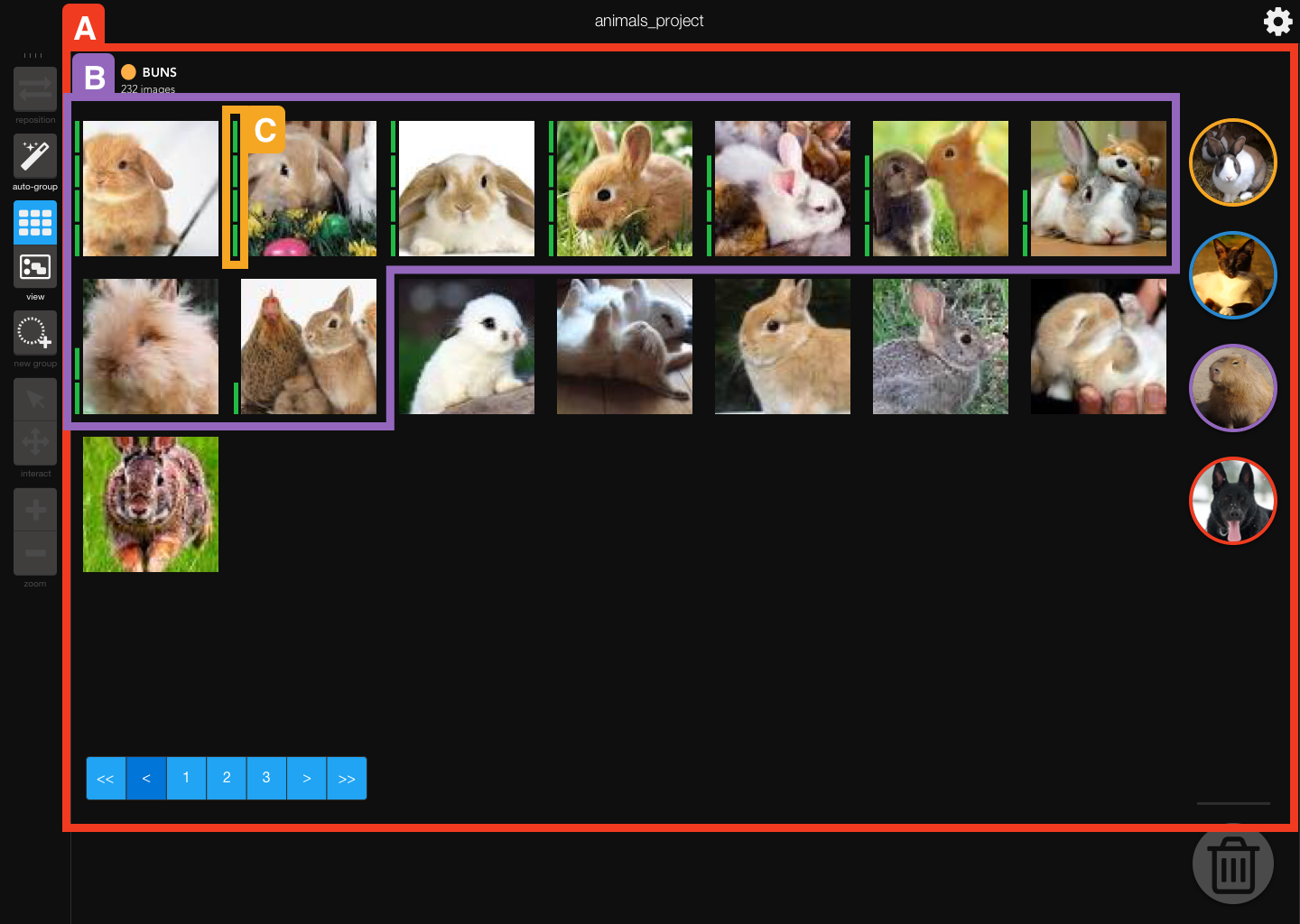}
   \caption{Sharkzor interface grid view. (A) The Sharkzor grid which supports the user's \textbf{triage} task. (B) Auto-grouped images. (C) Auto-group confidence visualization.}
   \label{fig:2}
\end{figure}

The \textbf{Triage} task involves taking the user’s initial query result image pile and turning it into an accurate visual representation of the user’s mental model. Functionality supporting this task includes pre-clustering and image interact modes. With images pre-clustered by visual similarity, the user can quickly grab groups of similar images from the canvas (Figure \ref{fig:1}A) and then organize them into groups (Figure \ref{fig:1}B) using grid view mode. The grid view mode (Figure \ref{fig:2}A) was designed specifically to support the triage task where on the initial canvas many images may be hidden behind others.

The \textbf{Organize} task revolves around user interaction with the canvas to create a visual model of image organization. Key functionality includes machine feedback via proximity (Figure \ref{fig:1}C) and grouping (Figure \ref{fig:1}B).

Once the other tasks have been completed within the user's workflow, Sharkzor can then expedite the user's workflow by \textbf{automating} common tasks. Sharkzor uses deep neural networks to automatically populate the visual representation of the user’s mental model. Functionality supporting this task includes auto-positioning and auto-grouping of images.

\subsection{Toward Understandable Machine Learning}
The Sharkzor system has affordances which provide transparency and increase trust in the underlying machine learning algorithms. These affordances are a step in the direction of making machine learning results less of a black box and more understandable to humans. These features include the (1) auto-group feedback, that images similar to those already grouped are added to existing groups (Figure \ref{fig:2}B) and (2) the confidence visualization (Figure \ref{fig:2}C) which provides a qualitative view of the algorithm's confidence in group assignment.

\section{Deep Learning}
\label{DL}
Sharkzor leverages multiple deep learning techniques to facilitate image identification and organization, including size-agnostic classification, pre-clustering and few-shot learning. These algorithms and methods are deployed into flexible and modular micro-services.

\subsection{Pre-Clustering}
Pre-clustering provides users with a starting arrangement of images from which to begin interacting.  A widely used algorithm exists precisely for this task in t-distributed stochastic neighbor embedding (tSNE) \cite{Maaten}. Unfortunately, tSNE becomes computationally prohibitive when the number of starting dimensions is greater than $O(10)$, as the case for image data. To this end, for Sharkzor a series of steps are taken to reduce an image's dimensionality before being passed to tSNE. The first step utilizes an autoencoder to compress images to 256 dimensions. The autoencoder is pre-trained on ImageNet images~\cite{deng2009imagenet}. After feature extraction using the autoencoder, principal component analysis is used to further reduce an image's dimensionality to eight dimensions. Finally, tSNE takes the image's eight dimensional representation and yields the desired two-dimensional embedding. The coordinates of the tSNE output are then utilized by the canvas for the initial image placement when the user starts Sharkzor.

\subsection{Few-Shot Learning}
To address the requirement of users being able to create arbitrary image-related mental models, we aren't able to use traditional multi-label classification techniques.
This is  because the user may be interested in clustering images into arbitrarily complex arrangements.
To make a  robust system that can adapt to user supplied groups, we leverage learning techniques requiring few trainging examples.

Few-shot learning techniques \cite{oneshot_siamese, oneshot_matchingnets} overcome the issue of requiring many hundreds of labeled examples as they classify images in relation to other images.
We accomplish this specifically by having our network learn a binary classification task to learn the probability that a reference image belongs in a class, where a class is a single image or a collection of images tagged by the user.

This provides key benefits in being flexible with respect to the number of groups.
If, for instance, we applied a final $n$-way classification layer on the few-shot model, such as in \cite{oneshot_matchingnets}, we would be locked into a fixed number of groups.
To support multi-group classification of images, we collect all of the output probabilities from comparing each ungrouped image to each user-provided group and then take the $argmax$ over the set of probabilities.
To allow pictures to remain ungrouped, we only assign a image to a group if  certainty exceeds a threshold.

\remove{\section{Architecture}
The Sharkzor micro-service architecture allows for arbitrary applications to integrate with and leverage complex machine learning techniques in an intuitive way to support image organization tasks.
The design of the backend architecture is specifically geared around being scalable, allowing for an arbitrary number of each service to run.}

\subsection{Training the neural networks}
Sharkzor leverages standard image datasets for training and performance assessment. These datasets include CalTech 256 \cite{caltech256}, CIFAR10 \& CIFAR100 \cite{cifar}, Visual Genome \cite{visgenome}, and Omniglot \cite{omniglot}.

It should be noted that the Sharkzor networks are trained to be explicitly class and data agnostic. Training the few-shot learning technique using common datasets  ensures that the system functions as expected. We can visually confirm results, and quickly experiment to benchmark things like performance versus class sizes.


\remove{\subsection{Image Service and UI Backend}

The image service serves as both the single point of access for images loading as well as a go-between to simplify communication between the UI and the learning service.
We have several bindings in this service to either load the previously mentioned benchmark datasets as well as user provided datasets.
The service holds access to all of the image data for usage within both the UI as well as the learning service. }

\subsection{Machine \& Deep Learning Service}

The machine \& deep learning micro-services are responsible for providing all functionality to requestors via the image service.
To operate, we initially leverage transfer learning by extracting features from ResNet \cite{resnet} for all of our images.
These features are then used for everything from training our exemplar regression model to training the few-shot model.
\remove{Additionally, we provide single-click tuning to the user-interface with the ability to use human-tagged groups from the UI to train the few-shot model.
To aid in rapid experimentation, we also provide functionality to reset tunable parameters back to their stock weights.}

\section{Conclusion}

In this work we describe our approach to accomplishing a deep learning assisted platform for visual triage, sort and summary of images, which encodes a user's mental model through human-in-the-loop interaction. We designed Sharkzor using a micro-services philosophy that aids users in instilling their mental model into the application through methods for pre-clustering, auto-grouping, and repositioning images using traditional machine learning, transfer learning and few-shot learning.

\section{Acknowledgements}
This work was funded the U.S. Government.

\bibliography{example_paper}
\bibliographystyle{icml2017}

\end{document}